\documentstyle[preprint,prb,aps]{revtex}

\begin{document}

\draft

\title{Finite temperature dynamic structure function of the free Bose
gas}

\author{F.Mazzanti$^{1,2}$ and A.Polls$^3$}
\address{$^1$ Institut f\"ur Theoretische Physik, 
	Johannes Kepler Universit\"at Linz, \\
	A-4040 Linz, Austria} 
\address{$^2$ Departament d'electr\`onica,
	Enginyeria i Arquitectura La Salle, \\
	Pg. Bonanova 8,  Universitat Ramon Llull,
	E-08022 Barcelona, Spain} 
\address{$^3$ Departament d'Estructura i Constituents
	de la Mat\`eria, \\
	Diagonal 645, Universitat de Barcelona,
	E-08028 Barcelona, Spain}

\maketitle

\begin{abstract}
A detailed calculation of the finite temperature dynamic structure
function of the free Bose gas is presented and discussed. After a
short derivation of the expressions describing the exact response
$S(q,\omega;T)$ and the Impulse Approximation (IA)
$S_{IA}(q,\omega;t)$, their main properties and their evolution
with $q$ and $T$ are analyzed. The lowest order energy
weighted sum rules of both $S(q,\omega;T)$ and $S_{IA}(q,\omega;T)$
are also derived and commented. Finally, the $q\to\infty$ asymptotic
behavior is analyzed and discussed in terms of scaling laws.
\end{abstract}

\maketitle

\pagebreak

  Neutron scattering on systems like pure $^4$He has proved to be
among the most fruitful methods to gather relevant information about
the fundamental properties of Bose condensed systems \cite{gly1}.
Although high momentum transfer neutron scattering is known to provide
much information about the structure of the Bose condensate, no
particular formalism has been able to accurately describe how
temperature influences the dynamic structure function $S(q,\omega)$
and how the temperature dependence of the condensation fraction value
affects $S(q,\omega)$ at high and low momentum transfers.  More
recently, the experimental discovery of Bose condensation in clouds of
alkali atoms has revived the interest in the analysis of free Bose
systems and the way in which Bose--Einstein condensation sets in
\cite{dalf,java}.  In previous works we presented detailed calculations of
the $T=0$ and $T>0$ dynamic structure function of the free Fermi gas,
their coherent and incoherent parts and the way in which they evolve
with momentum transfer \cite{mazz1,mazz2}. In this work we extend this
analysis to free Bose systems, focussing on the evolution of
$S(q,\omega)$ and its coherent and incoherent parts with $q$ and $T$,
the role of the Bose--Einstein condensate in the response and finally
on scaling laws.

The dynamic structure function $S(q,\omega;T)$ describes the way in
which a system of $N$ particles responds to a probe which produces a
density fluctuation modelized through the operator
$\rho_q=\sum_{j=1}^N e^{i{\bf q}\cdot{\bf r}_j}$.  at finite
temperature this is given by \cite{pin1,pin2,grif}
\begin{equation}
S(q,\omega;T) = \sum_{\{n,m\}}
\frac{1}{\mathcal{Z}}\,e^{-\beta \left(E_n-\mu N\right)}
\frac{1}{N}
\mid\! \left\langle m \mid \rho_q \mid n\right\rangle \!\mid^2
\delta\left( E_m - E_n - \omega \right) \ ,
\label{intro1}
\end{equation}

where $q$ and $\omega$ are the momentum and energy transferred
by the probe to the system.
${\mathcal{Z}}=\sum_{\{n\}} e^{-\beta(E_n-\mu N)}$
stands for the Grand Canonical partition function, while $\beta=1/k_BT$
is the inverse of the temperature and $\mu$ is the chemical potential.

The dynamic structure function is related by Fourier transformation to
the density--density response function $S(q,t)$ which may in turn be
separated in its coherent and incoherent parts
\begin{eqnarray}
S_{inc}(q,t;T)  = 
\sum_{\{n\}} \frac{1}{{\mathcal{Z}}}
e^{-\beta(E_n-\mu N)} \frac{1}{N} \sum_{j=1}^N
\left\langle n \mid e^{-i{\bf q}\cdot{\bf r}_j} e^{iHt}
e^{i{\bf q}\cdot{\bf r}_j} e^{-iHt}
\mid n \right\rangle
\label{intro2a} \\ [2mm]
S_{coh}(q,t;T)  =  
\sum_{\{n\}} \frac{1}{{\mathcal{Z}}}
e^{-\beta(E_n-\mu N)} \frac{1}{N} \sum_{i\neq j=1}^N
\left\langle n \mid e^{-i{\bf q}\cdot{\bf r}_i} e^{iHt}
e^{i{\bf q}\cdot{\bf r}_j} e^{-iHt}
\mid n \right\rangle \ ,
\label{intro2b}
\end{eqnarray}
$H$ and ${\bf r}_j$ being the Hamiltonian of the system and the
position operator of particle $j$, respectively. With these
definitions the incoherent response takes into account only the
diagonal terms of the density--density response function, while the
coherent response incorporates the contribution coming from all the
other terms in $S(q,t)$. Notice that these definitions are not unique
as in fact frequently $S_{coh}(q,t)$ is identified with $S(q,t)$ while
the last function in Eq.~(\ref{intro2b}) becomes simply the difference
between the new $S_{coh}(q,t)$ and $S_{inc}(q,t)$.

The dynamic structure function is also related to the imaginary part
of the dynamic susceptibility $\chi(q,\omega;T)$.  Taking into account
that $\chi(q,\omega;T)=-\pi(1-e^{-\beta\omega})
S(q,\omega;T)$~\cite{pin1,grif}, one finds
\begin{eqnarray}
\tilde{S}(\tilde{q},\tilde\omega;\tilde{T}) \equiv  
\epsilon_0 S(q,\omega;T) =
\frac{n_0(\tilde{T})}{1-e^{-\tilde\omega/\tilde{T}}}
\left[\delta\left(\tilde\omega-\tilde{q}^2\right) -
\delta\left(\tilde\omega+\tilde{q}^2\right)\right]
\nonumber \\ [2mm]
 -  \frac{\tilde{T}}{16\pi^2\tilde{q}}
\frac{1}{(1-e^{-\tilde\omega/\tilde{T}})}
\ln\left[\frac{1-z e^{-(\tilde\omega/\tilde{q}-q)^2/4\tilde{T}}}
{1-z e^{-(\tilde\omega/\tilde{q}+q)^2/4\tilde{T}}} \right] \ ,
\label{sqwb1}
\end{eqnarray}
that is written in terms of the dimensionless variables
$\tilde{q}=q/k_0$, $\tilde\omega=\omega/\epsilon_0$ and
$\tilde{T}=T/\epsilon_0$, where $k_0=\rho^{1/3}$ and
$\epsilon_0=k_0^2/2m$ define the momentum and energy scales.  Finally,
in Eq.~(\ref{sqwb1}) $z$ stands for the fugacity that is related to
the dimensionless chemical potential through
$z=e^{\tilde\mu/\tilde{T}}$, where $\tilde\mu=\mu/\epsilon_0$. This
last quantity can be derived recalling that at equilibrium particle
number conservation imposes
\begin{equation}
\int_0^\infty \frac{\epsilon^{1/2} d\epsilon}{z^{-1}
e^{\epsilon}-1}~=~ 
\frac{4\pi^2}{\tilde{T}^{3/2}}\left[1-n_0(\tilde{T})\right] \ ,
\label{sqwc1}
\end{equation}
where $n_0(\tilde{T})$ is the condensate fraction value at temperature
$\tilde{T}$ \cite{pat}. For the free Bose gas, at $T \leq T_c$
\begin{equation}
n_0(\tilde{T}) = 1-\left(\frac{\tilde{T}}{\tilde{T}_c}\right)^{3/2} \ ,
\label{sqwc1b}
\end{equation}
where $\tilde{T}_c\approx 6.63$ is the value of the dimensionless
Bose--Einstein transition temperature which in the current
dimensionless scheme becomes density and mass independent.

Notice that the first term in
Eq.~(\ref{sqwb1}) is proportional to the condensate fraction value
$n_0(\tilde T)$ and thus characteristic of Bose liquids: when the
system is given a net momentum transfer $\tilde q$, all particles in
the condensate respond equally to the perturbation and the response
peaks at a single frequency $\tilde\omega=\tilde q^2$. On the other
hand, the second term is the contribution of all the other particles
not laying in the condensate and is similar to the partticle--hole
contribution to the total response of Fermi systems.

In the free case and due to the absence of an interatomic potential,
the incoherent response coincides with the Impulse Approximation,
\begin{equation}
\tilde{S}_{inc}(\tilde{q},\tilde\omega; \tilde{T}) =
\tilde{S}_{IA}(\tilde{q},\tilde\omega; \tilde{T}) \equiv
\frac{1}{(2 \pi)^3} \int d\tilde{\bf k}\, n(\tilde{k})
\,\delta\!\left( (\tilde{\bf k}+\tilde{\bf q})^2 -
\tilde{k}^2 - \tilde\omega \right).
\label{sqwc2}
\end{equation}
Using the momentum distribution of the free Bose gas  one gets
the following result
\begin{equation}
\tilde{S}_{inc}(\tilde{q},\tilde\omega;\tilde{T})~=~ 
n_0(\tilde{T})\, \delta\!\left(\tilde \omega-\tilde{q}^2\right) -
\frac{\tilde{T}}{16\pi^2\tilde{q}} \ln\left[
1 - z e^{-(\tilde\omega/\tilde{q}-\tilde{q})^2/4\tilde{T}}
\right] \ ,
\label{sqwd1}
\end{equation}
while $\tilde{S}_{coh}(\tilde{q},\tilde\omega;\tilde{T})$ is the
difference between~(\ref{sqwb1}) and~(\ref{sqwd1}).

Below the Bose--Einstein transition temperature, where $z=1$, one can
introduce a new set of variables $Q=\tilde{q}/\tilde{T}^{1/2}$ and
$\nu=\tilde\omega/\tilde{T}$, such that the ratio of the
non--condensate contributions and $\tilde{T}^{1/2}$ becomes
temperature independent for all three responses when $Q$ and $\nu$ are
taken as new independent variables. Therefore, it is useful to
introduce three new structure functions that take this fact into
account,
\begin{eqnarray}
\hat{S}(Q,\nu) = \frac{1}{\sqrt{\tilde{T}}}
\tilde{S}_{nc}(\tilde{q},\tilde\omega;\tilde{T})
& \equiv & -\frac{1}{16\pi^2 Q} \frac{1}{(1-e^{-\nu})} \ln\left[
\frac{1-z e^{-(\nu/Q-Q)^2/4}}{1-z e^{-(\nu/Q+Q)^2/4}}
\right]
\label{sqwe1} \\ [1mm]
\hat{S}_{inc}(Q,\nu) = \frac{1}{\sqrt{\tilde{T}}}
\tilde{S}_{nc,inc}(\tilde{q},\tilde\omega;\tilde{T})
& \equiv & -\frac{1}{16\pi^2 Q} \ln\left[ 1 - z e^{-(\nu/Q-Q)^2/4} \right]
\label{sqw2} \\ [2mm]
\hat{S}_{coh}(Q,\nu) & \equiv &
\hat{S}(Q,\nu) -\hat{S}_{inc}(Q,\nu) \ ,
\label{sqwe3}
\end{eqnarray}
 where the subscript $nc$ indicates that only the non--condensate
contribution to the different responses has been taken into account. 

The total, coherent and incoherent non--condensate responses at
$\tilde{T}\!\leq\!\tilde{T}_c$ are plotted in Fig.~(\ref{fig-a}) for
$Q=0.7$, $Q=1.5$ and $Q=2$. Both the $\tilde{q}$ dependence and the
$\tilde{T}$ dependence of the responses are represented in there,
because $Q$ can be increased either by increasing $\tilde{q}$ at fixed
$\tilde{T}$ or by lowering $\tilde{T}$ at fixed $\tilde{q}$. In the
first case, the responses at different $Q$ can be directly compared,
while in the latter case one should bear in mind that
$\tilde{S}_{nc}(\tilde{q},\tilde\omega; \tilde{T}) = \tilde{T}^{1/2}
\hat{S}(Q,\nu)$.

As it is readily seen from Eq.~({\ref{sqwb1}), both the total and the
coherent responses of the free Bose gas presents logarithmic
singularities at $\tilde\omega=\tilde{q}^2$ and
$\tilde\omega=-\tilde{q}^2$, (that is, at $\nu=\pm Q^2$) when
$\tilde{T}$ is lower than the Bose--Einstein transition temperature,
as in this case $z=1$. On the other hand, at these temperatures
$\hat{S}_{inc}(Q,\nu)$ presents only one singularity located at
$\nu=Q^2$. In this way, the pole at $\nu=Q^2$ has contributions coming
from both the coherent and the incoherent responses, while the pole at
$\nu=-Q^2$ is of a completely coherent nature.  As $\tilde{T}$ grows
above the transition temperature, the chemical potential starts taking
negative values and the fugacity becomes smaller than $1$, thus moving
the poles from the real axis into the complex plane and hence removing
the previous singular behavior from all three responses.  In this
sense, the presence of these singularities in the non-condensate part
of the response may be considered as an indirect signature of the
existence of a Bose condensate.

At fixed $\tilde{T}$, the low $\tilde{q}$ response presents
nonvanishing contributions coming from both the coherent and the
incoherent parts of the response. Notice that, in contrast to what
happens in the free Fermi gas case~\cite{mazz2}, the low $\tilde{q}$'s
coherent response may actually overcome the incoherent one. As
$\tilde{q}$ is risen, the coherent response decays rapidly to zero and
thus the total response becomes mainly incoherent. As it is also
apparent from the figure, this is almost the case when $Q\approx 2$,
i.e., when $\tilde{q}\approx 2\tilde{T}^{1/2}$. On the other hand,
when $\tilde{q}$ remains fixed and $\tilde{T}$ is low, $Q$ becomes
large and hence $\hat{S}(Q,\nu)\approx\hat{S}_{inc}(Q,\nu)$, that is,
the total response becomes mainly incoherent. As the temperature is
risen, coherent contributions become more relevant and eventually,
depending on the value of $\tilde{q}$, at
$\tilde{T}\approx\tilde{q}^2$ both $\tilde{S}_{inc}$ and
$\tilde{S}_{coh}$ carry similar contributions to the total
response. Therefore the evolution with $\tilde{T}$ at fixed
$\tilde{q}$ can be understood as follows: at very low $\tilde{T}$, $Q$
is large and so the response behaves as if it were in the high
momentum transfer limit, where
$\hat{S}(Q,\nu)\approx\hat{S}_{inc}(Q,\nu)$ and
$\hat{S}_{coh}(Q,\nu)\to 0$. As the temperature rises, $Q$ shrinks and
takes its lowest value at $\tilde{T}=\tilde{T}_c$ before the Bose
condensate disapears.  When $\tilde{q}^2\gg\tilde{T}_c$, the response
still remains in the high momentum transfer limit, otherwise it enters
into the low momentum regime.

The behaviour of the three responses changes when the temperature
rises above the Bose--Einstein transition temperature. In fact, at
$\tilde{T}\!>\!\tilde{T}_c$ the chemical potential is strictly
negative and thus the fugacity $z$ is smaller than $1$.  Under such
circumstances, the poles of the responses that at
$\tilde{T}<\tilde{T}_c$ were at $\nu=\pm Q^2$ now move to $\nu=\pm Q^2
\pm i 2 Q\sqrt{-\ln z}$, and thus the singular points are removed from
the real $\nu$ axis, splitting each real pole into two complex
ones. When either $\tilde{q}$ or $\tilde{T}$ increases, the distance
from the poles to the real $\nu$ axis grows and the shape of the
response is smoothened.

This behaviour is shown in
Figs.~({\ref{fig-b}),~(\ref{fig-c}) and~(\ref{fig-d}) where the total,
coherent and incoherent responses are depicted for several values of
$Q$ in comparison with the classical response
$\hat{S}_{cl}(Q,\nu;\tilde{T})$ computed from a Maxwellian momentum
distribution
\begin{equation}
\hat{S}_{cl}(Q,\nu;\tilde{T}) = \frac{1}{2 Q\tilde T^{3/2}\sqrt{\pi}}
\, e^{-\frac{1}{4}\left(\frac{\nu}{Q}-Q\right)^2} \ ,
\label{sqwf1}
\end{equation}
which corresponds to the limiting case to which $\hat{S}(Q,\nu;\tilde{T})$
should approximate when the temperature increases.

As it can be seen from the figures, the way in which
$\hat{S}(Q,\nu;\tilde{T})$ approaches $\hat{S}_{inc}(Q,\nu;\tilde{T})$
(incoherent limit) and $\hat{S}_{cl}(Q,\nu;\tilde{T})$ (classical
limit) depends on the momentum transfer and on the temperature. At low
$\tilde{T}$ greater than $\tilde{T}_c$, the total response approaches
the incoherent one when $\tilde{q}$ grows. However, both
$\hat{S}(Q,\nu;\tilde{T})$ and $\hat{S}_{inc}(Q,\nu;\tilde{T})$ differ
from $\hat{S}_{cl}(Q,\nu;\tilde{T})$ because the momentum distribution
at those temperatures noticeably depart from the gaussian shape of the
classical $n(\tilde{k})$. When the temperature is risen, the
difference between the momentum distribution in the classical and
quantum quantum cases is reduced and so the high $\tilde{q}$ response
of the free Bose gas approaches the classical limit. Notice however
that the difference between $\hat{S}(Q,\nu;\tilde{T})$ and
$\hat{S}_{cl}(Q,\nu;\tilde{T})$ is still apparent at $\tilde{T}=20$,
which means that the classical limit is only well recovered at rather
high temperatures.

The evolution with $\tilde{T}$ of the different responses can also be
analyzed from the sum rules they satisfy. Sum rules are defined as the
different energy--weighted moments of the responses
\begin{eqnarray}
\tilde{m}_{inc,coh}^{(\alpha)}(\tilde{q})   \equiv  
\int_{-\infty}^\infty \tilde\omega^\alpha
\tilde{S}_{inc,coh}(\tilde{q}, \tilde\omega; \tilde{T})
\,d\tilde\omega
\nonumber \\ [2mm]
\tilde{m}^{(\alpha)}(\tilde{q}) \equiv
\tilde{m}_{inc}^{(\alpha)}(\tilde{q})   +  
\tilde{m}_{coh}^{(\alpha)}(\tilde{q}) =
\int_{-\infty}^\infty \tilde\omega^\alpha
\tilde{S}(\tilde{q}, \tilde\omega; \tilde{T}) \,d\tilde\omega \ .
\label{sr-b1}
\end{eqnarray}

The first lower orders sum rules can be easily deduced and yield the
following results
\begin{eqnarray}
\tilde{m}^{(0)}(\tilde{q}; \tilde{T}) & = & 
n_0(\tilde{T}) \left[
2\,n(\tilde{q}) + 1 \right] +
\tilde{m}^{(0)}_{nc}(\tilde{q}; \tilde{T}) 
= \tilde{S}(\tilde{q}; \tilde{T})
\label{sr-c1a} \\ [2mm]
\tilde{m}_{inc}^{(0)}(\tilde{q}; \tilde{T}) & = & 
n_0(\tilde{T}) + \tilde{m}_{inc,nc}^{(0)}(\tilde{q}; \tilde{T}) = 1
\label{sr-c1b} \\ [2mm]
\tilde{m}^{(0)}_{coh}(\tilde{q}; \tilde{T}) & = & 
2 n_0(\tilde{T})\,n(\tilde{q}) +
\tilde{m}_{coh,nc}^{(0)}(\tilde{q}; \tilde{T}) =
\tilde{S}(\tilde{q}; \tilde{T})-1
\label{sr-c1c} \\ [2mm]
\tilde{m}^{(1)}(\tilde{q}; \tilde{T}) & = & 
n_0(\tilde{T}) \tilde{q}^2 +
\tilde{m}^{(1)}_{nc}(\tilde{q}; \tilde{T}) = \tilde{q}^2
\label{sr-c1d} \\ [2mm]
\tilde{m}^{(1)}_{inc}(\tilde{q}; \tilde{T}) & = & 
n_0(\tilde{T}) \tilde{q}^2 +
\tilde{m}_{inc,nc}^{(1)}(\tilde{q}; \tilde{T}) = \tilde{q}^2
\label{sr-c1e} \\ [2mm]
\tilde{m}^{(1)}_{coh}(\tilde{q}; \tilde{T}) & = & 
\tilde{m}^{(1)}_{coh,nc}(\tilde{q}; \tilde{T}) = 0 \ ,
\label{sr-c1f}
\end{eqnarray}
where $\tilde{m}^{(\alpha)}_{nc}(\tilde{q})$ and
$\tilde{m}^{(\alpha)}_{(inc,coh),nc}(\tilde{q})$ refer to the
contribution to the sum rules coming from the integral of the
non--condensate parts of the different responses.

Eq.~(\ref{sr-c1a}) yields the values of the static structure factor
$\tilde{S}(\tilde{q}; \tilde{T})$ as obtained from the direct
integration of Eq.~(\ref{sqwb1}). Two different contributions appear
at temperatures below $\tilde{T}_c$. The first one results from the
condensate term in $\tilde{S}(\tilde{q}, \tilde\omega; \tilde{T})$ and
so is proportional to $n_0(\tilde{T})$. The contribution of this term
to the $\tilde{m}^{(0)}(\tilde{q}; \tilde{T})$ sum rule grows as
$n_0(\tilde{T})\left(\tilde{T}/\tilde{q}^2\right)$ at low momentum
transfer because at $\tilde{T}<\tilde{T}_c$ the low $\tilde{q}$ states
close to the condensate are populated according to a
$\tilde{T}/\tilde{q}^2$ law. Notice that this is an entirely coherent
effect as can be seen from Eqs.~(\ref{sr-c1b}) and~(\ref{sr-c1c}).
This behavour can be smeared out when particle interactions are
allowed to take place as happens in liquid $^4$He at low
temperatures. Also, as $\tilde{T}^{-1/2}\tilde{S}_{nc}(\tilde{q},
\tilde\omega; \tilde{T}) = \hat{S}_{nc}(Q,\nu;\tilde{T})$ is a
$\tilde{T}$--independent function of $Q$ and $\nu$ at
$\tilde{T}\leq\tilde{T}_c$, the zero order $\nu$--weighted moment of
$\hat{S}_{nc}(Q,\nu)$ yields an universal curve from where both the
evolution with $\tilde{q}$ and the evolution with $\tilde{T}$ of
$\tilde{S}_{nc}(\tilde{q}; \tilde{T}\leq\tilde{T}_c)$ can be
extracted. According to the previous definitions
$M^{(n)}_{nc}(Q)=\int \nu^n\hat{S}(Q,\nu;\tilde{T}) d\nu \equiv
\tilde{T}^{n+3/2} \tilde{m}_{nc}^{(n)}(\tilde{q}; \tilde{Y})$.

$M^{(0)}_{nc}(Q)$ is depicted in the upper plot of
Figure~(\ref{fig-e}).  At low $Q$, that is, at low momenta compared to
$\tilde{T}^{1/2}$, the divergences appearing in $\hat{S}(Q,\nu)$ at
$\nu=\pm Q^2$ take most of the strength and as a result
$M^{(0)}_{nc}(Q)$ increases. However, the singularities in
$\hat{S}(Q,\nu)$ at $\nu=\pm Q^2$ are of the logarithmic type and thus
their $\nu$--weighted integrals of the response give always finite
results consistent with Eqs.~(\ref{sr-c1d})--(\ref{sr-c1f}). On the
other hand, in the large $Q$ limit $M^{(0)}_{nc}(Q)$ approaches the
constant value $\tilde{T}_c^{-3/2}\approx 0.058$, a result that is
consistent with $\tilde{m}^{(0)}(\tilde{q},\tilde{T})$ going to
$1-n_0(\tilde{T})$ and the temperature dependence of the condensate
fraction value reported in Eq.~(\ref{sqwc1b}). When this value is
reached the total $\tilde{S}(\tilde{q},\tilde{T})$ equals $1$.  The
high $Q$ limit can be reached either rising $\tilde{q}$ at fixed
$\tilde{T}$ or lowering $\tilde{T}$ at fixed $\tilde{q}$. Therefore,
no matter what the value of $\tilde{q}$ is, at $\tilde{T}\!\to\!0$ the
static structure factor goes to $1$, a fact that is consistent with
$\tilde{S}(\tilde{q},\tilde\omega; \tilde{T}=0)$ being completely
incoherent and equal to $\delta\left(\tilde\omega-\tilde{q}^2\right)$.
When the temperature is risen, the total response receives
contributions from the non--condensate terms and thus
$\tilde{S}(\tilde{q}; \tilde{T})$ grows. Finally, when $\tilde{T}$
reaches the transition temperature $M^{(0)}_{nc}(Q)$ equals
$M^{(0)}(Q)$ but its value depends on the the momentum transfer. If
$\tilde{q}$ is still high compared to $\tilde{T}_c^{1/2}$, the
non--condensate part of the total response does not qualitatively
differ from the $\tilde{T}=0$ case and thus the the static structure
factor remains close to $1$. Otherwise, the response enters into the
low $Q$ regime as seen in Fig.~(\ref{fig-a}) and $\tilde{S}(\tilde{q};
\tilde{T})$ departs from unity. Taking into account that
$m^{(0)}(\tilde{q}; \tilde{T})=1$ and that the coherent response of
the free Bose gas is positive defined, the asymptotic value
$\tilde{S}(\tilde{q}; \tilde{T})=1$ is always reached from above.

When $\tilde{T}$ exceeds the Bose--Einstein transition temperature the
condensate fraction value decays to $0$ and the previous sum rules
coincide with their non--condensate parts. The evolution with
$\tilde{T}$ of $\tilde{S}(\tilde{q}; \tilde{T})$ is sketched in the
lower plot of Fig.~(\ref{fig-e}).

As the logarithmic singularities in the dynamic structure function are
smeared out at $\tilde{T}>\tilde{T}_c$, the limiting value
$\tilde{S}(\tilde{q}\!\to\!0; \tilde{T})$ now becomes finite. When
$\tilde{q}$ increases, $\tilde{S}(\tilde{q}; \tilde{T})$ decreases and
asymptotically approaches $1$ from above, thus indicating that the
total response is reaching the incoherent limit in which
$\tilde{S}(\tilde{q},\tilde\omega; \tilde{T})\!~\approx~\!
\tilde{S}_{inc}(\tilde{q},\tilde\omega; \tilde{T})$ and
$\tilde{S}_{coh}(\tilde{q},\tilde\omega; \tilde{T})\!~\!\approx~0$.
In the high $\tilde{T}$ limit $\tilde{S}(\tilde{q};
\tilde{T})\!\approx\!1$ at all $\tilde{q}$'s and this reveals that the
classical limit has almost been reached, as
$\tilde{S}_{cl}(\tilde{q};\tilde{T})$ is known to be equal to $1$ for
all $\tilde{q}$. Moreover this limit is consistently reached when
$\tilde{T}\!\to\!\infty$, as in fact it is straightforwardly seen from
the very definition of $\tilde{T}$ that this can be achieved either
increasing the temperature at fixed density or reducing the density at
fixed temperature.

Another interesting feature of the high momentum transfer behaviour of
the response is the  scaling property, as it is well known that
when $\tilde{q}\!\to\!\infty$ the dynamic structure function does not
depend anymore on $\tilde{q}$ and $\tilde\omega$ separately but only
through the West scaling variable
$\tilde{y}=(\tilde\omega/\tilde{q}-\tilde{q})/2$\cite{west}.

In the free Bose gas case scaling may be reached in two different ways
depending on the temperature. At $\tilde{T}\!\leq\!\tilde{T}_c$ this
can be done introducing a Compton profile 
$\hat{J}^-_{nc}(Q,Y;\tilde{T})$ as follows
\begin{equation}
\hat{J}^-_{nc}(Q,Y;\tilde{T}) = 2 Q\,\hat{S}(Q,\nu;\tilde{T})
\label{j-b1}
\end{equation}
where $Y=\left(\nu/Q-Q\right)/2$. With this definition,
the non--condensate contribution to the incoherent response becomes
temperature and momentum independent
\begin{equation}
\hat J^-_{inc,nc}(Y) = -\frac{1}{8\pi^2} \ln\left[1-e^{-Y^2} \right] \ ,
\label{j-c1}
\end{equation}
and so becomes an universal function valid for all values of $Y$ as
long as $\tilde{T}\leq\tilde{T}_c$. The non--condensate contribution
to the total response reaches the incoherent limit at high momentum
transfers, so $\hat{J}^-_{nc}(Q,Y;\tilde{T})$ approaches
$\hat{J}^-_{nc,inc}(Y)$ when $Q\!\to\!\infty$. This is indeed the
case as shown in Fig.~(\ref{fig-f}). As can be seen, already at $Q=2$
the incoherent limit seems to be almost reached, which means that at
any temperature below $\tilde{T}_c$ scaling is well recovered when the
transferred momentum is larger than
$\tilde{q}=2\sqrt{\tilde T_c}\!\approx\!5$.

Scaling can also be recovered at temperatures above the Bose--Einstein
transition temperature. However in that case the following alternative
definition of the Compton profile may be used
\begin{equation}
\hat{J}^+(Q,Y;\tilde{T}) = 2Q\,\tilde{T}^{3/2} \hat{S}(Q,\nu;\tilde{T}) \ .
\label{j-d1}
\end{equation}
With this definition, the classical response becomes $Q$ and
$\tilde{T}$--independent according to the expression
\begin{equation}
\hat{J}^+_{cl}(Y) = \frac{1}{\sqrt{\pi}}\,e^{-Y^2} \ ,
\label{j-d2}
\end{equation}
and therefore this is the limiting case to which the response of the
free Bose gas tends when the temperature is risen. The response
$\hat{J}^+(Q,Y;\tilde{T})$ and the incoherent response
$\hat{J}^+_{inc}(Q,Y;\tilde{T})$ at $Q=1$ are compared at different
$\tilde{T}$'s with $\hat{J}_{cl}^{+}(Y)$ in Fig.~(\ref{fig-g}). Both
functions approach the classical behaviour when the temperature is
risen, even though it is remarkable how far should the temperature go
in order to reach that limit. In any case and as expected, the
incoherent response approaches faster $\hat{J}^+_{cl}(Y)$, a fact that
can be understood recalling that the total response contains also
coherent contributions that do not scale in the same variables. In
this way, the different velocity at which both functions approach
$\hat{J}^+_{cl}(Y)$ gives a measure of how fast the coherent response
goes to zero.  Finally the main reason why the scaling in
Eq.~(\ref{j-d1}) is proportional to $\tilde{T}^{3/2}$ may be
understood recalling that in the classical regime $z$ is proportional
to $\tilde{T}^{3/2}$, so that in the end both dependencies cancel and
$\hat{J}^+_{cl}(Y)$ becomes $\tilde{T}$--independent. This explains
why the scaling law at $\tilde{T}>\tilde{T}_c$ and
$\tilde{T}\!\leq\!\tilde{T}_c$ should be different, as in the latter
case the fugacity vanishes and therefore it can not introduce any
explicit dependence on $\tilde{T}$ as happens at high temperatures.

In summary, the dynamic structure function of the free Bose gas at
finite temperatures is shown to contain both coherent and incoherent
contributions. At low temperatures compared to the Bose--Einstein
transition temperature $T_c$ , the total response is mainly incoherent
and tends to the limiting incoherent value $\delta(\omega-q^2/2m)$.
When the temperature is risen, coherent contributions presenting
logarithmic singularities at $\omega=\pm q^2/2m$ appear, and they
actually overcome the contribution of the incoherent response when
$T\approx T_c$.  When $T$ is increased above $T_c$, the logarithmic
divergences disappear but a visible signature of their presence below
$T_c$ still remain in both $S_{inc}$ and $S_{coh}$.  When $T\gg T_c$
the coherent response cancels while the incoherent and total responses
approach the classical limit.

The incoherent limit is always recovered when the transferred momentum
is high, while the classical limit is only reached at high $T$. Below
$T_c$, the presence of singularities in the response is also reflected
in their lowest order energy--weighted sum rules. Finally, in the high
$q$ limit the total response can be written in the form of a Compton
profile that scales in the West scaling variable $Y$. The temperature
dependence can also be removed from the response by using an
approppriate redefinition of $Y$, even though the transformation is
different at $T<T_c$ and at $T>T_c$ due to the different temperature
dependence of the chemical potential.  Despite the simplicity of the
system analyzed and the fact that the excitation spectrum of an
interacting bose system will be different the results described in
this work are expected to enlight some aspects of the finite
temperature response of both weakly and strong interacting Bose
systems.

\acknowledgements

This work has been  partially supported by the Austrian Science Fund under
grant No. P12832-TPH,   DGICYT (Spain) grant No. PB95-1249 and 
the program SGR98-11 from {\em Generalitat de Catalunya}.


\begin{figure}
\caption{ Non-condensate contributions to the total ($\hat{S}(Q,\nu)$,
solid lines), coherent ($\hat{S}_{coh}(Q,\nu)$, dashed lines) and
incoherent ($\hat{S}_{inc}(Q,\nu)$, dotted lines), for $Q=0.7,1.5$ and
$2$, at any temperature below $\tilde T_c$.}
\label{fig-a}
\end{figure}

\begin{figure}
\caption{Temperature dependence above $\tilde T_c$ of the total ($\hat
S(Q,\nu;\tilde{T})$, solid line), coherent ($\hat
S_{coh}(Q,\nu;\tilde{T})$, dashed lines) and incoherent ($\hat
S_{inc}(Q,\nu;\tilde{T})$, dotted line), in all cases at $Q=0.7$. The
classical limit $\hat S_{cl}(Q,\nu;\tilde{T})$ is also shown at each
temperature (dot-dahsed lines).}
\label{fig-b}
\end{figure}

\begin{figure}
\caption{Same plots as in Fig.~(\ref{fig-b}) at $Q=1.5$.}
\label{fig-c}
\end{figure}

\begin{figure}
\caption{Same plots in Fig.~(\ref{fig-b}) at $Q=2$.}
\label{fig-d}
\end{figure}

\begin{figure}
\caption{Contribution of the non--condensate part of the response to
the zero order sum rule. Upper plot: $M_{nc}^{(0)}(Q)$ at
$\tilde{T}\leq\tilde{T}_c$. Lower plot:
$\tilde{m}^{(0)}(\tilde{q},\tilde {T})$ at $\tilde{T}>\tilde{T}_c$,  for
$\tilde{T}=7$ (solid line), $\tilde{T}=8$ (dotted line),
$\tilde{T}=10$ (short-dashed line), and $\tilde{T}=20$ (long-dashed
line).}
\label{fig-e}
\end{figure}

\begin{figure}
\caption{$Q$ dependence of the Compton profile $\hat{J}^-_{nc}(Q,Y)$
at $\tilde T < \tilde T_c$ compared with the universal
$\hat{J}_{inc,nc}^-(Y)$ .}
\label{fig-f}
\end{figure}

\begin{figure}
\caption{ Temperature dependence in the range $\tilde T > \tilde T_c$
at fixed $Q=1$ of total $\hat{J}^+(Q,Y;\tilde{T})$ (left hand
side) and the incoherent $\hat{J}^+_{inc}(Q,Y;\tilde{T})$ (right
hand side), compared to the universal $\hat{J}^+_{cl}(Y)$(solid line).}
\label{fig-g}
\end{figure}

\pagebreak





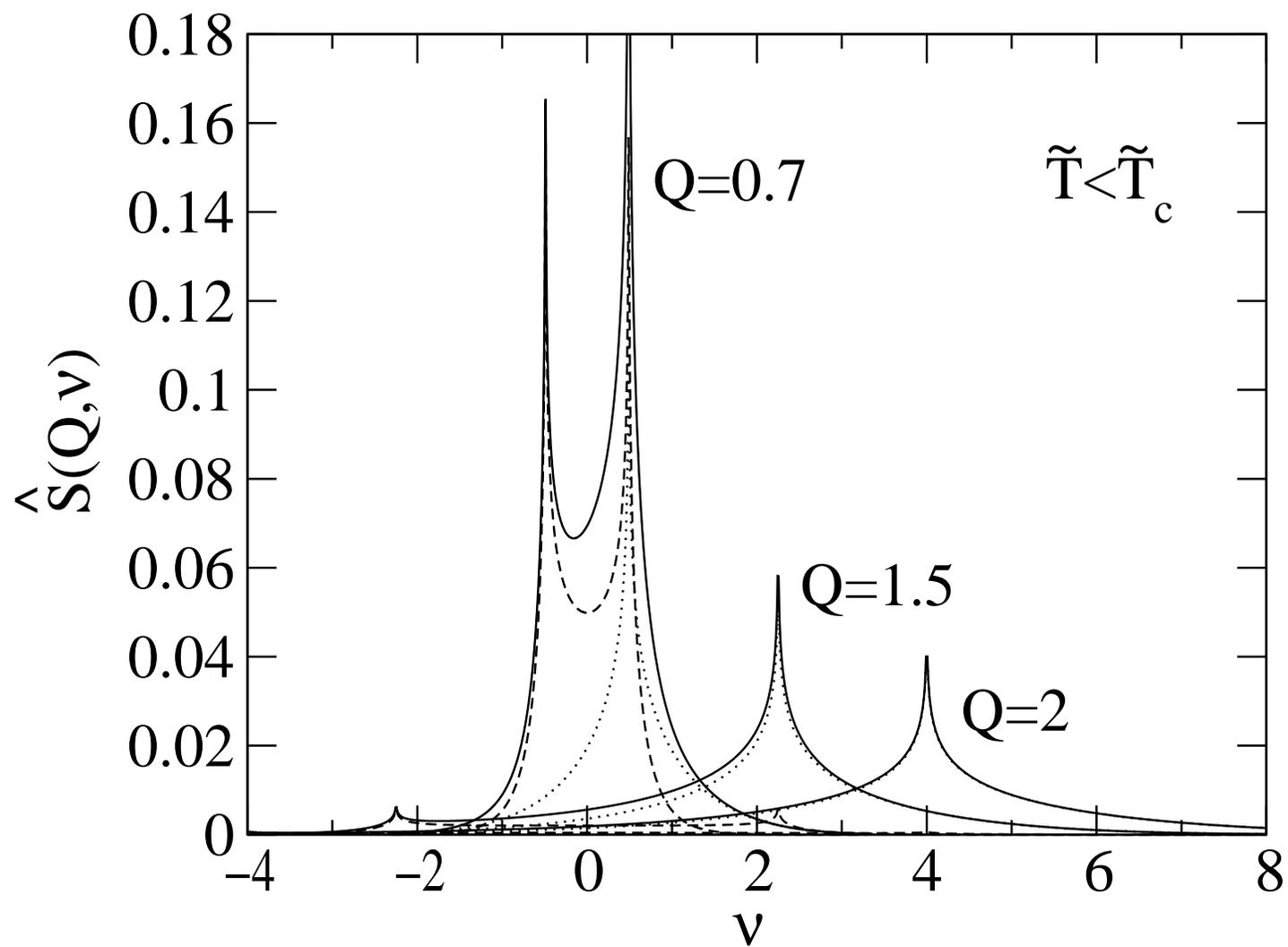
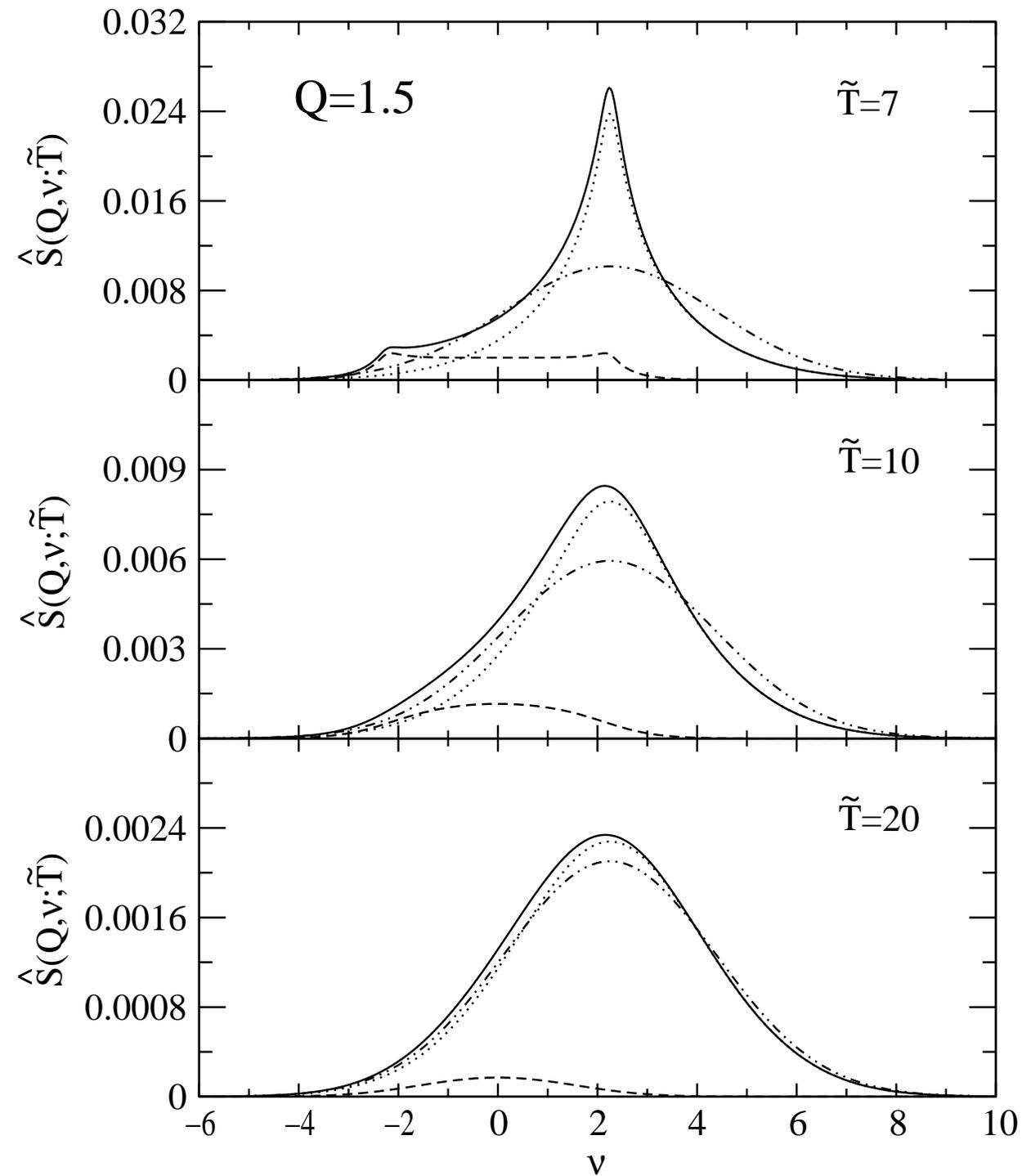
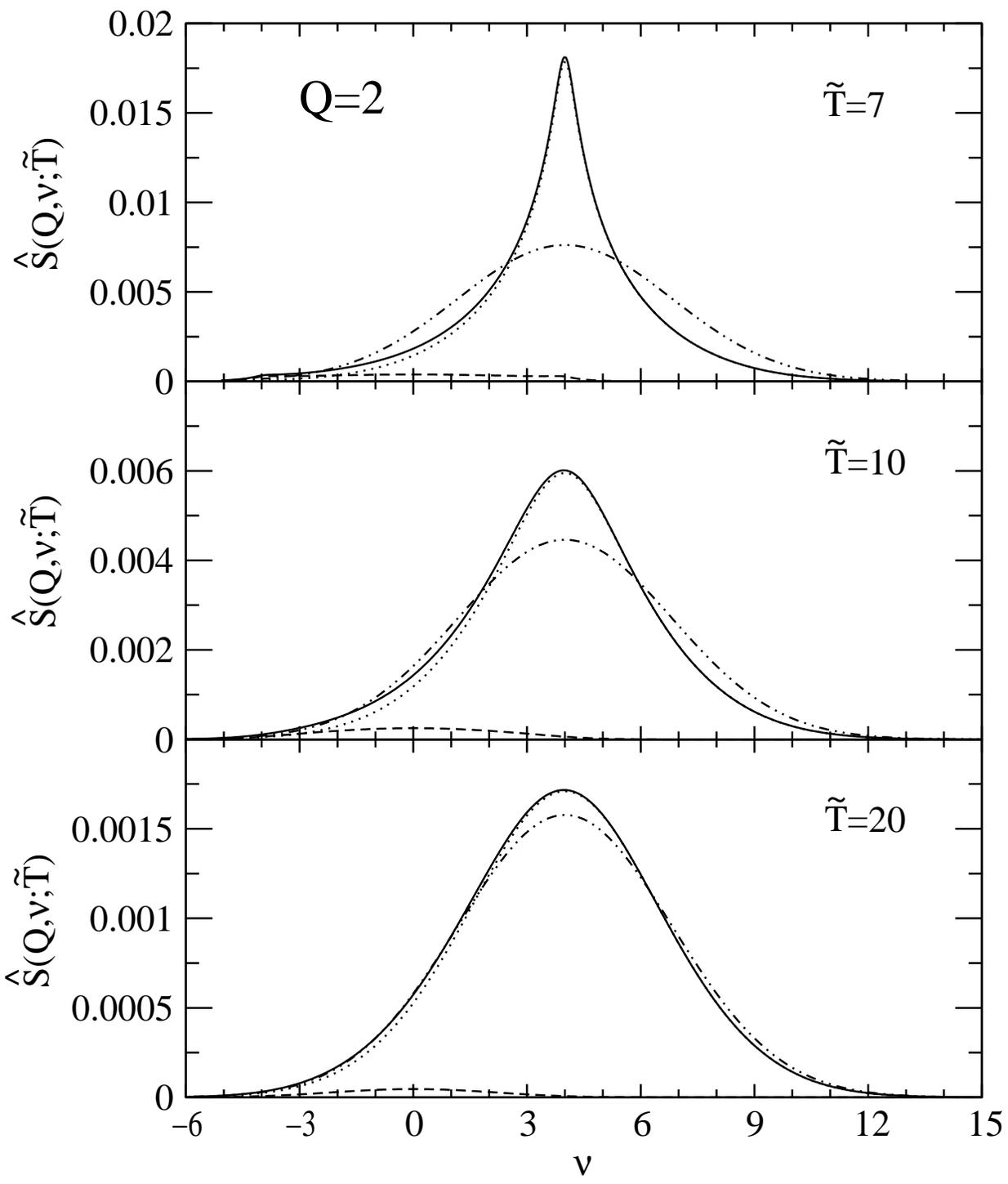
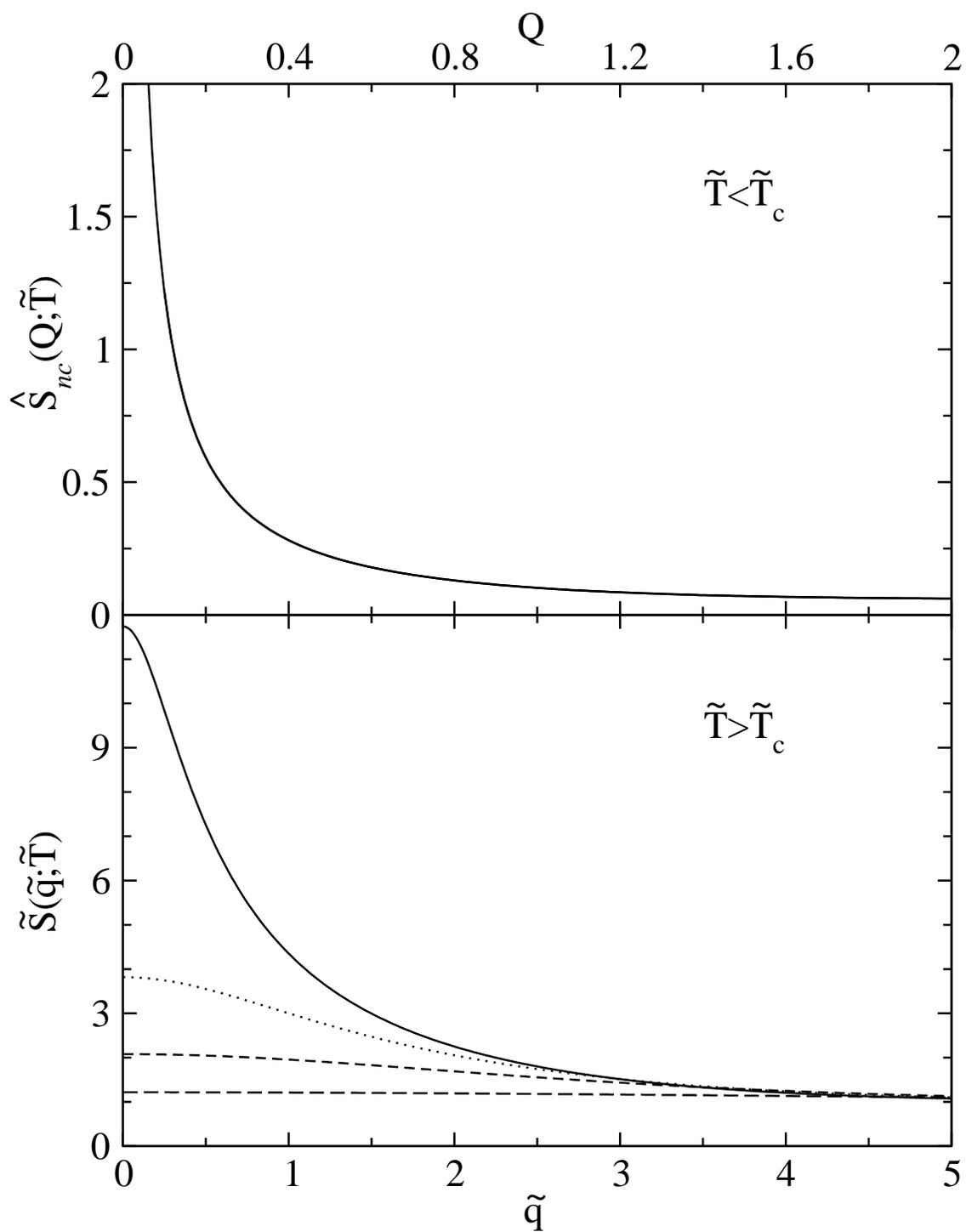
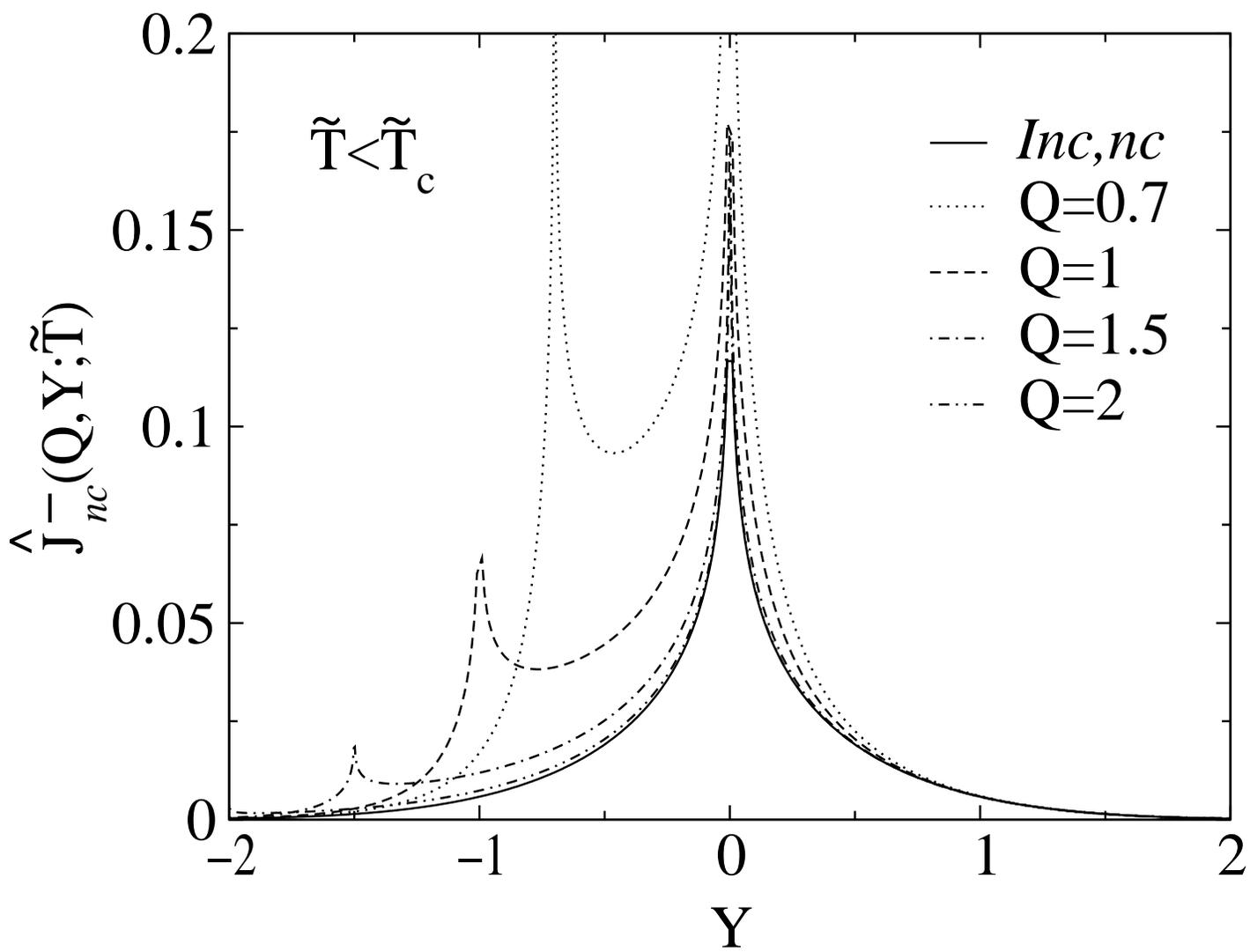
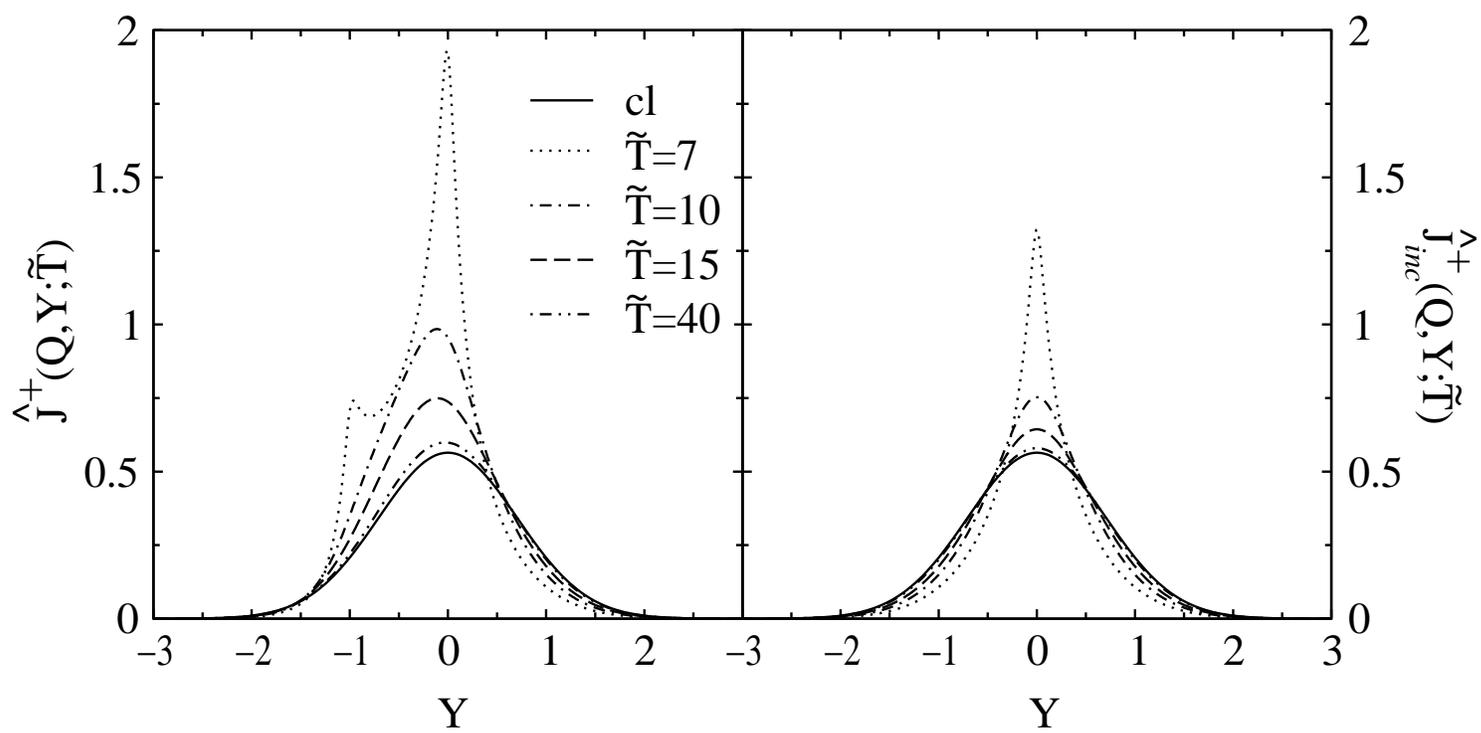
\end{document}